\documentclass[american,aps,pra,reprint,superscriptaddress,10pt]{revtex4-1}
\usepackage[T1]{fontenc}
\usepackage[sc]{mathpazo}
\usepackage{amsmath}
\usepackage{amssymb}
\usepackage{enumerate}
\usepackage{amsthm}% theorems and proofs
\usepackage{color}

\usepackage{amsfonts,mathrsfs}
\usepackage[style]{fncychap}
\usepackage{graphicx} % For including graphics N.B. pdftex graphics driver
\usepackage{geometry} % allows easy specifying of the page layout
\usepackage{eepic}
\usepackage{ifthen}
\newboolean{ElectronicVersion}
\setboolean{ElectronicVersion}{true} % CHANGE THIS AS REQUIRED

\usepackage[T1]{fontenc}
\usepackage[sc]{mathpazo}
\usepackage{mathdots}
\usepackage{amsmath}
\usepackage{amssymb}
\usepackage{enumerate}
\usepackage{amsthm}% theorems and proofs
\usepackage{amsfonts,mathrsfs}
\usepackage[style]{fncychap}
\usepackage{graphicx} % For including graphics N.B. pdftex graphics driver
\usepackage{geometry} % allows easy specifying of the page layout
\usepackage{eepic}
\usepackage{ifthen}
\usepackage{cases}

\usepackage{amsfonts,mathrsfs}
\usepackage[style]{fncychap}
\usepackage{graphicx} % For including graphics N.B. pdftex graphics driver
\usepackage{geometry} % allows easy specifying of the page layout
\usepackage{eepic}
\usepackage{ifthen}
\usepackage{cases}

\usepackage{hyperref}
\hypersetup{pdfpagemode=UseNone}

\def\be{\begin{equation}}
\def\ee{\end{equation}}
\def\bea{\begin{eqnarray*}}
\def\eea{\end{eqnarray*}}

\newenvironment{mylist}[1]{\begin{list}{}{
    \setlength{\leftmargin}{#1}
    \setlength{\rightmargin}{0mm}
    \setlength{\labelsep}{2mm}
    \setlength{\labelwidth}{8mm}
    \setlength{\itemsep}{0mm}}}
    {\end{list}}

\newcommand{\ket}[1]{|#1\rangle}

\newcommand{\braket}[1]{\langle#1\rangle}

\def\H{\textsf{H}}

\theoremstyle{definition}

\numberwithin{equation}{section}
\def\be{\begin{equation}}
\def\ee{\end{equation}}
\def\bea{\begin{equation*}}
\def\eea{\end{equation*}}
\def\bna{\begin{eqnarray*}}
\def\ena{\end{eqnarray*}}
\def\bn{\begin{eqnarray}}
\def\en{\end{eqnarray}}
\def\bpm{\begin{pmatrix}}
\def\epm{\end{pmatrix}}

\newcounter{questionnumber}

\usepackage{chngcntr}
\counterwithout{equation}{section}

\begin{document}\pagestyle{empty}

\title{To quantify the difference of $\eta$-inner products in $\cal PT$-symmetric theory}

 \author{Minyi Huang}
 \email{11335001@zju.edu.cn}
 \email{hmyzd2011@126.com}
 \affiliation{Department of Mathematical Sciences, Zhejiang Sci Tech University, Hangzhou 310007, PR~China}

 \author{Guijun Zhang}

\affiliation{School of Mathematical Sciences, Zhejiang University, Hangzhou 310027, PR~China}

\begin{abstract}
In this paper, we consider a typical continuous two dimensional $\cal PT$-symmetric Hamiltonian and propose two different
approaches to quantitatively show the difference between the $\eta$-inner products. Despite the continuity of Hamiltonian, the
$\eta$-inner product is not continuous in some sense. It is shown that the difference between the $\eta$-inner products of
broken and unbroken $\cal PT$-symmetry is lower bounded. Moreover, such a property can lead to some uncertainty relation.
\end{abstract}

\maketitle

\section{Introduction}
The $\cal PT$-symmetric quantum theory, which was established by C. M. Bender and his colleagues, has been stimulating our understanding of many interesting mathematical and physical problems since its inception \cite{bender1998real,guo2009observation,ruter2010observation}.
%In its development, this has found lots of applications in many branches of physics \cite{guo2009observation,ruter2010observation}.
Recently, owing to the explorations of non-Hermitian dynamics and topology \cite{ashida2020nonhermitian}, there is also a growing research interest in the theory of
$\cal PT$-symmetry.

Initially, $\cal PT$-symmetric quantum theory was established as an alternative description framework of quantum systems \cite{bender2007making}.
In such a framework, a closed $\cal PT$-symmetric system should give a unitary time evolution in certain sense. However,
the non-Hermiticity of a $\cal PT$-symmetric operator generally prevents from a unitary evolution in the usual Dirac
inner product. To this end, the $\cal CPT$ inner product was introduced, under which the $\cal PT$-symmetric systems can
evolve unitarily. The concept of $\cal CPT$ inner product was later generalized to a more general setting,
namely the $\eta$-inner product. Utilizing the $\eta$-inner product, $\cal PT$-symmetric quantum theory was generalized
to the pseudo-Hermitian theory \cite{mostafazadeh2010pseudo}. Despite a theoretical concept, $\eta$-inner product was shown to be tightly linked
to the experimental realization of $\cal PT$-symmetric systems \cite{gunther2008naimark,PhysRevLett.119.190401,huang2018embedding}. Indeed, it is not only the key concept of pseudo-Hermitian
theory but also a mark of distinction between different types of $\cal PT$-symmetric systems. Usually, $\cal PT$-symmetric systems can be classified into two types, the unbroken and broken $\cal PT$-symmetric systems.
In the case of unbroken $\cal PT$-symmetry, the $\eta$-inner product can be either definite or indefinite. On the other hand,
the $\eta$-inner product of broken $\cal PT$-symmetry can only be indefinite \cite{huang2018embedding}.

This paper focuses on the discussion of $\eta$-inner product,
especially on the distinction between the $\eta$-inner products of broken and unbroken $\cal PT$-symmetry. We
consider a typical continuous two dimensional $\cal PT$-symmetric Hamiltonian and propose two different approaches to
quantify the difference between the $\eta$-inner products. Despite the continuity of Hamiltonian, the $\eta$-inner
product is not continuous in some sense. It is shown that
the difference between the $\eta$-inner products of broken and unbroken $\cal PT$-symmetry is lower bounded. Moreover, such a property can lead to an uncertainty relation.

The remaining  part of this paper is organized as follows. In section II, we briefly introduce some basic notions of $\cal PT$-
symmetry. In section III, we give a method to quantify the difference between $\eta$-inner products and obtain some formal uncertainty relation. In section IV, we discuss another method to measure the difference between metric operators. In section V, some discussions are made.

%=============================================================================%

\section{Preliminaries}
\subsection{Basic notions of $\cal PT$-symmetric systems}

Some basic notions of finite dimensional $\cal PT$-symmetric systems are briefly introduced as follows.

A parity operator $\cal P$ is a linear operator such that ${\cal P}^2={\cal I}$, where ${\cal I}$ is the identity operator on $\mathbb{C}^d$.

A time reversal operator $\cal T$ is an anti-linear operator such that ${\cal
T}^2={\cal I}$. Moreover, the parity and time reversal operators should commute, that is,
${\cal PT}={\cal TP}$.

A linear operator $\cal H$ on $\mathbb{C}^d$ is said to be $\cal
PT$-symmetric if ${\cal H}{\cal PT}={\cal PT}{\cal H}$.

In finite dimensional case, a linear operator corresponds uniquely to a matrix and an anti-linear operator
corresponds to the composition of a matrix and
the complex conjugation \cite{uhlmann16}.
Let $A$ be a matrix, with $\overline{A}$ the complex conjugation of $A$ and $A^\dag$ the transpose of $\overline{A}$.
Let $P$, $T$ and $H$  be the matrices of ${\cal P}$, ${\cal T}$ and
$\cal H$, respectively. Then the definition conditions of $\cal P$, $\cal T$, $\cal H$ are
$P^2=T\overline{T}=I$, $PT=T\overline{P}$ and $HPT=PT\overline{H}$.

By considering the spectral property of $H$, $\cal PT$-symmetric systems can be classified into two classes:

A $\cal PT$-symmetric operator $\cal H$ is said to be unbroken if $H$ is similar to a real diagonal matrix;

A $\cal PT$-symmetric operator $\cal H$ is said to be broken if $H$ cannot be diagonalised or has complex eigenvalues.

\subsection{Two dimensional model}
A typical two dimensional $\cal PT$-symmetric Hamiltonian was given by C. M. Bender \cite{bender2007making}. In
this paper, we consider a special case of that example,
\be H(\theta)=E_0I_2+s\begin{bmatrix} i\sin\theta&1\\1&-i\sin\theta\end{bmatrix}.\label{HG}\ee
By taking ${\cal P}=\sigma_x$ and $\cal T$ the complex conjugation, one can directly verify that $H(\theta)$
is $\cal PT$-symmetric.
Apparently, this Hamiltonian is continuous with respect to $\theta$.
The eigenvalues are $\lambda_\pm=E_0\pm s\cos\theta$ and
 the corresponding eigenvectors are
\begin{eqnarray*}
\psi_+(\theta)=
\frac{1}{\sqrt{2}}
\begin{bmatrix}
e^{i\frac{\theta}{2}} &\\
e^{-i\frac{\theta}{2}} &
\end{bmatrix},
\psi_-(\theta)=\frac{1}{\sqrt{2}}
\begin{bmatrix}
 e^{-i\frac{\theta}{2}} \\
-e^{i\frac{\theta}{2}}
\end{bmatrix}.
\end{eqnarray*}
$\theta=\pm\frac{\pi}{2}$ is the exceptional point, at which $H$ cannot be diagonalized in general.
 In this case, the two eigenvectors coalesce.
 When $\alpha$ takes other values, $\cal PT$-symmetry is unbroken.

\subsection{The concepts of metric operator and dilation}

 $\cal PT$-symmetric systems are often viewed as effective models in the sense of open systems, and experimenters can use large
 Hermitian systems to simulate $\cal PT$-symmetric systems. The simulation of $\cal PT$-symmetric systems is tightly related to
 the mathematical concept of dilation.

Assume that $H$ is a $\cal PT$-symmetric Hamiltonian, which is either time dependent or independent.
By dilating $H$ to a Hermitian Hamiltonian, we mean that one can find some Hermitian operator
$\hat{H}=\begin{bmatrix}H_1&H_2\\H_2^\dag & H_4\end{bmatrix}$ and $\tau$ such that for any vector $\psi$, %$\begin{bmatrix}
%\psi\\
%\tau\psi
%\end{bmatrix}$ be a state.

\be
\begin{bmatrix}
i\psi'\\
i(\tau\psi)'
\end{bmatrix}
=
\begin{bmatrix}
H_1&H_2\\
H_2^\dag&H_4
\end{bmatrix}
\begin{bmatrix}
\psi\\
\tau\psi
\end{bmatrix}
=
\begin{bmatrix}
H\psi\\
i(\tau\psi)'
\end{bmatrix}
.\label{e}
\ee
For the first component, we see that
\[
i\psi'=H\psi.
\]
According to the Sch\"{o}dinger equation,
 the effect of $H$ will be realized.

In fact, if we denote \be\eta=I+\tau^\dag\tau,\label{condition}\ee then the dilation problem is equivalent to
finding $\eta$ such that
\be i\eta'=H^\dag\eta-\eta H.\label{eta}\ee
Usually, the (invertible) Hermitian operator $\eta$ in the above equation is called the metric operator \cite{mostafazadeh2010pseudo,zhang2019time}.
In particular, when $\eta$ is time independent, Eq. (\ref{eta}) reduces to
\be
H^\dag\eta=\eta H.\label{e1}
\ee
Note that Eq. (\ref{condition}) implies that $\eta>0$, that is, $\eta$ is positive definite.
It can be proved that the solutions to Eqs. (\ref{condition}) and (\ref{e1}) exist if and only if $H$ is unbroken $\cal PT$-symmetric
\cite{huang2018embedding}. On the other hand, if we relax the condition only considering Eq. (\ref{e1}), then the $\eta$ also exists for
any broken $\cal PT$-symmetric Hamiltonian. However, in this case, $\eta$ is indefinite (has negative eigenvalues and cannot be used for dilation.
 A detailed discussion of the time independent dilation problem
can be found in in \cite{gunther2008naimark,PhysRevLett.119.190401,huang2018embedding}.
To simulate the broken $\cal PT$-symmetric systems, researchers usually consider the time dependent case, namely Eq. (\ref{eta}). In this case, there are also interesting results \cite{wu2019observation,zhang2019time}. In this paper, we only consider the time independent case and the metric operators always refer to the operators satisfying Eq. (\ref{e1}).

The above discussion of (time independent) dilation contains Eq. (\ref{HG}) as a special case and experimenters have realized such a dilation in lab \cite{gunther2008naimark,tang2016experimental}. It should also be mentioned that the effect of a $\cal PT$-symmetric
system is in the subspace. Hence we may discuss the dilation efficiency. Eq. (\ref{e}) shows that for $\begin{bmatrix}
\psi\\
\tau\psi
\end{bmatrix}$, its first component $\psi$ is effectively subject to the $\cal PT$-symmetric Hamiltonian $H$. For any $\ket{\psi}$,
the transition probability from  $\begin{bmatrix}
\psi\\
\tau\psi
\end{bmatrix}$ to $\ket{\psi}$ is $|\frac{\braket{\psi|\psi}}{\braket{\psi|I+\tau^2|\psi}}|=|\frac{\braket{\psi|\psi}}{\braket{\psi|\eta|\psi}}|$.
Then the dilation efficiency $E_d$ is characterized by the minimal transition probability over all $\ket{\psi}$,
\be E_d=p_{min}=\min\limits_{\ket{\psi}}|\frac{\braket{\psi|\psi}}{\braket{\psi|\eta|\psi}}|=\frac{1}{\lambda_{max}},
\label{E}\ee
where $\lambda_{max}$ is the
maximal eigenvalue of the metric operator $\eta$.

\subsection{The structure of finite dimensional $\cal PT$-symmetric systems}
Finite dimensional $\cal PT$-symmetric Hamiltonians, as well as their metric operators, have some special structures  \cite{gohberg1983matrices2,huang2018embedding}. As an example, for the Hamiltonian $H$ in Eq. (\ref{HG}) and its certain metric operator $\eta$,
there exist an invertible matrix $\Psi$ such that
\be
H=\Psi\Lambda \Psi^{-1} \text{and}~\eta=(\Psi^{-1})^\dag D \Psi^{-1},
\ee
where $\Lambda$ and $D$ are the canonical forms of $H$ and $\eta$, respectively. To be precise, when $\cal PT$-symmetry is unbroken,
$\Lambda$ is a diagonal matrix whose diagonal entries are the eigenvalues of $H$, $D$ is a diagonal matrix whose diagonal entries are
$\pm 1$. % and $\Psi$ consists of two eigenvectors of $H$.
When  $\cal PT$-symmetry is broken, $\Lambda$ is the two dimensional Jordan canonical form of $H$, namely
$\begin{bmatrix}E_0&1\\0&E_0\end{bmatrix}$, and $D=\sigma_x$ (Pauli matrix) \cite{gohberg1983matrices2,huang2018embedding}.
Some observations can be made based on the above conclusion. Firstly, for Eq. (\ref{HG}), only unbroken $\cal PT$-symmetric Hamiltonians
can have positive definite metric operators. Secondly, one can use the matrix $\Psi$ to construct the metric operators. In particular, when $\cal PT$-symmetry is unbroken, the two column vectors of matrix $\Psi$ are just the eigenvectors of $H$.

\section{the metric operator and an Uncertainty relation}

In Eq. (\ref{HG}), the Hamiltonian $H$ is a continuous function of $\theta$.
Moreover, it is unbroken $\cal PT$-symmetric except for $\theta=\pm\frac{\pi}{2}$. Now we consider the following problem: Can one find some metric operator $\eta$ such that it is positive definite in the unbroken $\cal PT$-symmetric phase and is continuous at the exceptional point?

The answer is negative in general.  In fact, if $\eta>0$ for any $\theta\neq \pm\frac{\pi}{2}$, then by the continuity of
$\eta$, $\eta(\pm\frac{\pi}{2})\geqslant 0$, which contradicts with the indefiniteness.
This can also be seen through direct calculation.

Denote
$
\eta(\theta)=
\begin{bmatrix}
\eta_{11}(\theta)&\eta_{12}(\theta)\\
\eta_{21}(\theta)&\eta_{22}(\theta)
\end{bmatrix},
$
where $\eta_{21}^\dag(\theta)=\eta_{12}(\theta)$.
Now Eq. (\ref{e1}) reduces to the following
\begin{eqnarray*}
&&\begin{bmatrix}
-is\sin\theta & s\\
s & is\sin\theta
\end{bmatrix}
\begin{bmatrix}
\eta_{11}(\theta) & \eta_{12}(\theta)\\
\eta_{21}(\theta) & \eta_{22}(\theta)
\end{bmatrix}\\
&=&
\begin{bmatrix}
\eta_{11}(\theta) & \eta_{12}(\theta)\\
\eta_{21}(\theta) & \eta_{22}(\theta)
\end{bmatrix}
\begin{bmatrix}
is\sin\theta & s\\
s & -is\sin\theta
\end{bmatrix}.
\end{eqnarray*}
It follows that
\begin{eqnarray*}
&&s(\eta_{12}^\dag(\theta)-\eta_{12}(\theta))=i2s\sin\theta \eta_{11}(\theta),\\
&&s(\eta_{11}(\theta)-\eta_{22}(\theta))=0,\\
&&s(\eta_{11}(\theta)-\eta_{22}(\theta))=0,\\
&&s(\eta_{12}^\dag(\theta)-\eta_{12}(\theta))=i2s\sin\theta \eta_{22}(\theta).
\end{eqnarray*}
Hence
\be
\eta(\theta)
=\eta_{11}(\theta)\begin{bmatrix}
1&a(\theta)-i\sin\theta\\
a(\theta)+i\sin\theta&1
\end{bmatrix},\label{eta1}
\ee
where $a(\theta)$ is a real number. In addition, the two eigenvalue of $\eta(\theta)$ are
\begin{eqnarray*}
\lambda_+(\theta)=\eta_{11}(\theta)(\frac{1+\sqrt{a^2(\theta)+\sin^2\theta}}{2}),\\
\lambda_-(\theta)=\eta_{11}(\theta)(\frac{1-\sqrt{a^2(\theta)+\sin^2\theta}}{2}).
\end{eqnarray*}

From Eq. (\ref{eta1}), we find that when the Hamiltonian is at the exceptional point, that is $\theta= \pm\frac{\pi}{2}$,
the two eigenvalues of $\eta(\pm\frac{\pi}{2})$ are $\eta_{11}(\pm\frac{\pi}{2})(\frac{1\pm\sqrt{a^2(\pm\frac{\pi}{2})+1}}{2})$.
Note that the metric operator is invertible, which implies that $a(\pm\frac{\pi}{2})\neq 0$. Hence one of the eigenvalues of
$\eta(\pm\frac{\pi}{2})$ must be negative. On the other hand, the metric operator is positive definite when $\theta\neq \pm\frac{\pi}{2}$, which shows that $\eta(\theta)$ is not continuous.

Now we use the $l_1$ norm to measure whether the metric operators of unbroken and broken $\cal PT$-symmetry are close, under the
assumption that $\eta$ is positive definite when $\cal PT$-symmetry is unbroken. The definition of $l_1$ norm is
\be
\|\eta(\theta)\|_{l_1}=\sum_{ij}|\eta_{ij}(\theta)|.
\ee
Since $a(\theta)$ and $a(\frac{\pi}{2})$ are real numbers, the $l_1$ distance of $\eta(\theta)$ and $\eta(\frac{\pi}{2})$ can be written as
\begin{eqnarray*}
&&\Delta_1(\theta)=\|\eta(\theta)-\eta(\frac{\pi}{2})\|_{l_1}=2|\eta_{11}(\theta)-\eta_{11}(\frac{\pi}{2})|+\\
&&2|\eta_{11}(\theta)a(\theta)-\eta_{11}(\frac{\pi}{2})a(\frac{\pi}{2})-i[\eta_{11}(\theta)\sin\theta-\eta_{11}(\frac{\pi}{2})]|.
\end{eqnarray*}
%Note that , hence we have
%\begin{eqnarray*}
%&&2|\eta_{11}(\theta)a(\theta)-\eta_{11}(\frac{\pi}{2})a(\frac{\pi}{2})-i(\eta_{11}(\theta)\sin\theta-\eta_{11}(\frac{\pi}{2}))|\\
%&&\geqslant 2|\eta_{11}(\theta)\sin\theta-\eta_{11}(\frac{\pi}{2})|.
%\end{eqnarray*}
It follows that
\begin{eqnarray*}
\Delta_1(\theta)&\geqslant&2|\eta_{11}(\theta)-\eta_{11}(\frac{\pi}{2})|+2|\eta_{11}(\theta)\sin\theta-\eta_{11}(\frac{\pi}{2})|\\
&\geqslant& 2\eta_{11}(\theta)(1-\sin\theta).
\end{eqnarray*}

As was mentioned, $\lambda_\pm>0$ when $\theta\neq \pm\frac{\pi}{2}$.
Similar to Eq. (\ref{E}), one can formally define \[p_-(\theta)=\frac{1}{\lambda_-(\theta)}.\]
Here $p_-(\theta)$ cannot be interpreted as some transition probability, since $p_-(\theta)$ may be larger than one.
 Direct calculations show that
 \begin{eqnarray*}
 \nonumber &&\Delta_1(\theta) p_-(\theta)=|\frac{4(1-\sin\theta)}{1-\sqrt{a^2(\theta)+\sin^2\theta}}|\\
 && \geqslant |\frac{4(1-\sin\theta)}{1-|\sin\theta|}|\geqslant 4.
 \end{eqnarray*}
That is, one can obtain some uncertainty relation
 \be
\Delta_1(\theta) p_-(\theta)\geqslant 4.\label{ed1}
 \ee

In particular, when $\lambda_-(\theta)>1$,
 one can further obtain some relation
between $\Delta_1$ and the dilation efficiency,
 \begin{eqnarray}
\frac{\Delta_1(\theta)}{E_d(\theta)}\geqslant \frac{4}{p_-(\theta)}\cdot\lambda_+(\theta)\geqslant4.\label{ed2}
 \end{eqnarray}

\section{ Another way to measure the difference between metric operators}
We propose another way to see how close the unbroken $\eta$-inner product is to the broken
case. As was mentioned in Section. II. D, the metric operators can be constructed by utilizing the eigenvectors $\psi_+(\theta)$ and
$\psi_-(\theta)$. However, at the exceptional point, the two eigenvectors coalesce, which brings some inconvenience. To this end, we consider the following two states,
\begin{eqnarray*}
\psi_+(\theta)=
\frac{1}{\sqrt{2}}
\begin{bmatrix}
e^{i\frac{\theta}{2}} &\\
e^{-i\frac{\theta}{2}} &
\end{bmatrix},
\psi_-(\theta')=\frac{1}{\sqrt{2}}
\begin{bmatrix}
 e^{-i\frac{\theta'}{2}} \\
-e^{i\frac{\theta'}{2}}
\end{bmatrix}.
\end{eqnarray*}
Apparently, $\psi_+(\theta)$ is the eigenstate of $H(\theta)$ and $\psi_-(\theta')$ is the eigenstate of $H(\theta')$.
Now take the matrix
\begin{eqnarray*}
\Psi(\theta,\theta')=[\psi_+(\theta), \psi_-(\theta')]=
\frac{1}{\sqrt{2}}
\begin{bmatrix}
e^{i\frac{\theta}{2}} & e^{-i\frac{\theta'}{2}} \\
e^{-i\frac{\theta}{2}} & -e^{i\frac{\theta'}{2}}
\end{bmatrix}.
\end{eqnarray*}
Direct calculations show that
\begin{eqnarray*}
&&\Psi(\theta,\theta')^{-1}=
\frac{1}{\sqrt{2}\cos\frac{\theta+\theta'}{2}}
\begin{bmatrix}
e^{i\frac{\theta'}{2}} & e^{-i\frac{\theta'}{2}} \\
e^{-i\frac{\theta}{2}} & -e^{i\frac{\theta}{2}}
\end{bmatrix},\\
&&(\Psi(\theta,\theta')^{-1})^\dag=
\frac{1}{\sqrt{2}\cos\frac{\theta+\theta'}{2}}
\begin{bmatrix}
e^{-i\frac{\theta'}{2}} & e^{i\frac{\theta}{2}} \\
e^{i\frac{\theta'}{2}} & -e^{-i\frac{\theta}{2}}
\end{bmatrix}.
\end{eqnarray*}
Moreover,
\begin{eqnarray*}
&&H(\theta)[\psi_+(\theta),\psi_-(\theta')]\\
&=&[\psi_+(\theta),\psi_-(\theta')]
\begin{bmatrix}
E_0+s\cos\theta & 2si \sin\frac{\theta-\theta'}{2}\\
0 & E_0-s\cos\theta
\end{bmatrix}.
\end{eqnarray*}
Denote
\[
\Lambda(\theta,\theta')=\begin{bmatrix}
b_1(\theta,\theta')&c(\theta,\theta')i\\
0&b_2(\theta,\theta')
\end{bmatrix},
\]
where
\begin{eqnarray}
&&b_1(\theta,\theta')=E_0+s\cos\theta,\\
&&b_2(\theta,\theta')=E_0-s\cos\theta,\\
&&c(\theta,\theta')=2s \sin\frac{\theta-\theta'}{2}.
\end{eqnarray}
Now we have
\begin{eqnarray*}
&&H(\theta)=\Psi(\theta,\theta')\Lambda(\theta,\theta')\Psi^{-1}(\theta,\theta'),\\ &&\Psi^\dag(\theta,\theta')\eta(\theta,\theta')\Psi(\theta,\theta')=D(\theta,\theta').
\end{eqnarray*}
In this case, $H(\theta)^\dag\eta(\theta,\theta')=\eta(\theta,\theta') H(\theta)$ is equivalent to
\[
\Lambda(\theta,\theta')^\dag D(\theta,\theta')=D(\theta,\theta')\Lambda(\theta,\theta').
\]
It follows that
\[
\begin{bmatrix}
b_1&0\\
-ci&b_2
\end{bmatrix}
\begin{bmatrix}
d_{11}&d_{12}\\
d_{21}&d_{22}
\end{bmatrix}
=
\begin{bmatrix}
d_{11}&d_{12}\\
d_{21}&d_{22}
\end{bmatrix}
\begin{bmatrix}
b_1&ci\\
0&b_2
\end{bmatrix},
\]
where $d_{ij}$ are the entries of $D$ and $d_{12}=d_{21}^\dag$. Also note that we have omitted the symbol of $(\theta,\theta')$ in
the above equations for simplicity.
Direct calculations show that
\begin{eqnarray*}
&&b_1d_{11}=d_{11}b_1,\\
&&b_1d_{12}=d_{11}ci+d_{12}b_2,\\
&&-cd_{11}i+b_2d_{21}=d_{21}b_1,\\
&&-cd_{12}i+b_2d_{22}=cd_{21}i+d_{22}b_2.
\end{eqnarray*}
The above equations reduce to the following conditions,
\begin{eqnarray}
&&(b_1-b_2)d_{12}=d_{11}ci,\label{d1}\\
&&(b_1-b_2)d_{21}=-cd_{11}i,\label{d2}\\
&&c(d_{21}+d_{12})i=0.\label{d3}
\end{eqnarray}
When $\theta\neq \pm\frac{\pi}{2}$, then
\begin{eqnarray}
&&d_{12}=\frac{d_{11} \sin\frac{\theta-\theta'}{2}}{\cos\theta}i,\label{12}\\
&&d_{21}=-\frac{d_{11} \sin\frac{\theta-\theta'}{2}}{\cos\theta}i.\label{21}
\end{eqnarray}
For the exceptional point, that is when $\theta=\pm\frac{\pi}{2}$, Eqs. (\ref{d1})-(\ref{d3}) imply that $d_{11}=0$ and $d_{12}$ is
an imaginary number.
In this case, $|\frac{d_{11}}{d_{12}}|=0$. Denote $\delta=|\frac{d_{22}(\frac{\pi}{2})}{d_{12}(\frac{\pi}{2})}|$.
Now define the following measure
\be
\Delta_2(\theta,\theta')=|\frac{d_{11}}{d_{12}}(\theta,\theta')|+|\frac{d_{22}}{d_{12}}(\theta,\theta')-\delta|.
\ee
Apparently, a smaller value of $\Delta_2$ means that the $\eta$ inner product structure is closer to that of
broken $\cal PT$-symmetry. Now we show that there is a positive lower bound for $\Delta_2$, if $\eta$ is positive definite in the
unbroken $\cal PT$-symmetric case. Note that $\eta(\theta,\theta')>0$ is equivalent to $D(\theta,\theta')>0$, which implies that
\be
|d_{11}d_{22}|>|d_{12}|^2.\label{d}
\ee
 Consequently, we have
\[
|\frac{d_{22}}{d_{12}}|>|\frac{d_{12}}{d_{11}}|.
\]
Since $\Delta_2\geqslant 0$, we need only to show that $0$ is not the infimum of $\Delta_2$.
In fact, if $|\frac{d_{11}}{d_{12}}(\theta,\theta')|$ is sufficiently small, then $|\frac{d_{22}}{d_{12}}(\theta,\theta')|$, and thus the $\Delta_2$ will be large enough. Hence $0$ is not the infimum and $\Delta_2$ has a positive lower bound.

In particular, one can consider the case $\delta=0$, which is tightly related to the canonical form of metric operators \cite{gohberg1983matrices2}.
According to Eqs. (\ref{12}) and (\ref{21}),
\[
|\frac{d_{11}}{d_{12}}|=|\frac{d_{11}}{d_{21}}|=|\frac{\cos\theta}{\sin\frac{\theta-\theta'}{2}}|.
\]
Denote $d_{22}=td_{11}$. Eqs. (\ref{12}), (\ref{21}) and (\ref{d}) imply that
\[
t\geqslant (\frac{\sin\frac{\theta-\theta'}{2}}{\cos\theta})^2.
\]
Hence we know
\be
\Delta_2\geqslant|\frac{\cos\theta}{\sin\frac{\theta-\theta'}{2}}|+|\frac{\sin\frac{\theta-\theta'}{2}}{\cos\theta}|\geqslant 2.
\ee

\section{Discussions}
Firstly, the two ways to construct metric operators in section III and IV
are equivalent. To see this, not that we can take $\theta'=\theta-\pi$ in Eqs. (\ref{12}) and (\ref{21}).
Direct calculations show that
\begin{eqnarray}
\nonumber&&\eta(\theta)=\frac{1}{2\sin^2\theta}\\
\nonumber&&\begin{bmatrix}
d_{22}-d_{11}&-d_{11}e^{-i\theta}-d_{22}e^{i\theta}+\frac{2d_{11}}{\cos\theta}\\
-d_{11}e^{i\theta}-d_{22}e^{-i\theta}+\frac{2d_{11}}{\cos\theta}&d_{22}-d_{11}
\end{bmatrix}.\\
\label{eta2}
\end{eqnarray}
Apparently, Eq. (\ref{eta1}) and Eq. (\ref{eta2}) have the same form. At the exceptional point, Eqs. (\ref{d1})-(\ref{d3}) imply that $d_{11}=0$. 
Furthermore, for simplicity we take $\theta=\frac{\pi}{2}$, $\theta'=-\frac{\pi}{2}$. Direct calculations show that
\be
\eta(\frac{\pi}{2})=\frac{1}{2}\begin{bmatrix}
d_{22}&(-2d_{12}i)-d_{22}i\\
(-2d_{12}i)+d_{22}i &d_{22}
\end{bmatrix},
\ee
which has the same form as Eq. (\ref{eta1}). It implies that by fixing the value of $\theta'$, one can obtain all the
metric operators and the constructions of metric operators in section III and IV are equivalent.

Secondly, note that in our discussions, the Hamiltonian is time independent, hence the metric operator is not continuous at the exceptional point.
However, when the Hamiltonian is time dependent, its metric operators are often continuous \cite{wu2019observation}.

At the end, the reason why researchers prefer positive metric operators is that they can be used to dilate
$\cal PT$-symmetric Hamiltonians to large Hermitian ones and realize effective $\cal PT$-symmetric systems. This explains why
we consider the condition that the metric operator is positive in the unbroken $\cal PT$-symmetric phase. Eq. (\ref{ed2}) shows that when $\Delta_1$ is small, $E_d(\theta)$ is also small. It means that when the metric operator is closer to the broken
$\cal PT$-symmetry, an unbroken $\cal PT$-symmetric system is more difficult to be effectively realized.

%\section{Metric operator}
%\be
%\eta=(\Psi^{-1})^\dag\Psi^{-1}=\frac{1}{\cos^2\theta}
%\begin{bmatrix}
%1& -i\sin\theta\\
%i\sin\theta & 1
%\end{bmatrix}
%\ee
%The two eigenvalues of the metric operator are $\frac{1\pm\sin\theta}{\cos^2\theta}$. When $t\rightarrow \pm\frac{\pi}{2}$, at least one of the two eigenvalues tends to infinity. The

\section*{Acknowledgement}
This work is partially supported by the National Natural Science Foundation of China (11901526), the Postdoctoral Science Foundation of China (2020M680074) and the Science Foundation of Zhejiang Sci-Tech University (19062117-Y).
%

%\bibliographystyle{apsrev4-1}
%\bibliography{index}

\begin{thebibliography}{14}%
\makeatletter
\providecommand \@ifxundefined [1]{%
 \@ifx{#1\undefined}
}%
\providecommand \@ifnum [1]{%
 \ifnum #1\expandafter \@firstoftwo
 \else \expandafter \@secondoftwo
 \fi
}%
\providecommand \@ifx [1]{%
 \ifx #1\expandafter \@firstoftwo
 \else \expandafter \@secondoftwo
 \fi
}%
\providecommand \natexlab [1]{#1}%
\providecommand \enquote  [1]{``#1''}%
\providecommand \bibnamefont  [1]{#1}%
\providecommand \bibfnamefont [1]{#1}%
\providecommand \citenamefont [1]{#1}%
\providecommand \href@noop [0]{\@secondoftwo}%
\providecommand \href [0]{\begingroup \@sanitize@url \@href}%
\providecommand \@href[1]{\@@startlink{#1}\@@href}%
\providecommand \@@href[1]{\endgroup#1\@@endlink}%
\providecommand \@sanitize@url [0]{\catcode `\\12\catcode `\$12\catcode
  `\&12\catcode `\#12\catcode `\^12\catcode `\_12\catcode `\%12\relax}%
\providecommand \@@startlink[1]{}%
\providecommand \@@endlink[0]{}%
\providecommand \url  [0]{\begingroup\@sanitize@url \@url }%
\providecommand \@url [1]{\endgroup\@href {#1}{\urlprefix }}%
\providecommand \urlprefix  [0]{URL }%
\providecommand \Eprint [0]{\href }%
\providecommand \doibase [0]{http://dx.doi.org/}%
\providecommand \selectlanguage [0]{\@gobble}%
\providecommand \bibinfo  [0]{\@secondoftwo}%
\providecommand \bibfield  [0]{\@secondoftwo}%
\providecommand \translation [1]{[#1]}%
\providecommand \BibitemOpen [0]{}%
\providecommand \bibitemStop [0]{}%
\providecommand \bibitemNoStop [0]{.\EOS\space}%
\providecommand \EOS [0]{\spacefactor3000\relax}%
\providecommand \BibitemShut  [1]{\csname bibitem#1\endcsname}%
\let\auto@bib@innerbib\@empty
%</preamble>
\bibitem [{\citenamefont {Bender}\ and\ \citenamefont
  {Boettcher}(1998)}]{bender1998real}%
  \BibitemOpen
  \bibfield  {author} {\bibinfo {author} {\bibfnamefont {C.~M.}\ \bibnamefont
  {Bender}}\ and\ \bibinfo {author} {\bibfnamefont {S.}~\bibnamefont
  {Boettcher}},\ }\href@noop {} {\bibfield  {journal} {\bibinfo  {journal}
  {Phys. Rev. Lett.}\ }\textbf {\bibinfo {volume} {80}},\ \bibinfo {pages}
  {5243} (\bibinfo {year} {1998})}\BibitemShut {NoStop}%
\bibitem [{\citenamefont {Guo}\ \emph {et~al.}(2009)\citenamefont {Guo},
  \citenamefont {Salamo}, \citenamefont {Duchesne}, \citenamefont {Morandotti},
  \citenamefont {Volatier-Ravat}, \citenamefont {Aimez}, \citenamefont
  {Siviloglou},\ and\ \citenamefont {Christodoulides}}]{guo2009observation}%
  \BibitemOpen
  \bibfield  {author} {\bibinfo {author} {\bibfnamefont {A.}~\bibnamefont
  {Guo}}, \bibinfo {author} {\bibfnamefont {G.}~\bibnamefont {Salamo}},
  \bibinfo {author} {\bibfnamefont {D.}~\bibnamefont {Duchesne}}, \bibinfo
  {author} {\bibfnamefont {R.}~\bibnamefont {Morandotti}}, \bibinfo {author}
  {\bibfnamefont {M.}~\bibnamefont {Volatier-Ravat}}, \bibinfo {author}
  {\bibfnamefont {V.}~\bibnamefont {Aimez}}, \bibinfo {author} {\bibfnamefont
  {G.}~\bibnamefont {Siviloglou}}, \ and\ \bibinfo {author} {\bibfnamefont
  {D.}~\bibnamefont {Christodoulides}},\ }\href@noop {} {\bibfield  {journal}
  {\bibinfo  {journal} {Phys. Rev. Lett.}\ }\textbf {\bibinfo {volume} {103}},\
  \bibinfo {pages} {093902} (\bibinfo {year} {2009})}\BibitemShut {NoStop}%
\bibitem [{\citenamefont {R{\"u}ter}\ \emph {et~al.}(2010)\citenamefont
  {R{\"u}ter}, \citenamefont {Makris}, \citenamefont {El-Ganainy},
  \citenamefont {Christodoulides}, \citenamefont {Segev},\ and\ \citenamefont
  {Kip}}]{ruter2010observation}%
  \BibitemOpen
  \bibfield  {author} {\bibinfo {author} {\bibfnamefont {C.~E.}\ \bibnamefont
  {R{\"u}ter}}, \bibinfo {author} {\bibfnamefont {K.~G.}\ \bibnamefont
  {Makris}}, \bibinfo {author} {\bibfnamefont {R.}~\bibnamefont {El-Ganainy}},
  \bibinfo {author} {\bibfnamefont {D.~N.}\ \bibnamefont {Christodoulides}},
  \bibinfo {author} {\bibfnamefont {M.}~\bibnamefont {Segev}}, \ and\ \bibinfo
  {author} {\bibfnamefont {D.}~\bibnamefont {Kip}},\ }\href@noop {} {\bibfield
  {journal} {\bibinfo  {journal} {Nat. Phys.}\ }\textbf {\bibinfo {volume}
  {6}},\ \bibinfo {pages} {192} (\bibinfo {year} {2010})}\BibitemShut {NoStop}%
\bibitem [{\citenamefont {Ashida}\ \emph {et~al.}(2020)\citenamefont {Ashida},
  \citenamefont {Gong},\ and\ \citenamefont {Ueda}}]{ashida2020nonhermitian}%
  \BibitemOpen
  \bibfield  {author} {\bibinfo {author} {\bibfnamefont {Y.}~\bibnamefont
  {Ashida}}, \bibinfo {author} {\bibfnamefont {Z.}~\bibnamefont {Gong}}, \ and\
  \bibinfo {author} {\bibfnamefont {M.}~\bibnamefont {Ueda}},\ }\href@noop {}
  {\enquote {\bibinfo {title} {Non-hermitian physics},}\ } (\bibinfo {year}
  {2020}),\ \Eprint {http://arxiv.org/abs/2006.01837} {arXiv:2006.01837
  [cond-mat.mes-hall]} \BibitemShut {NoStop}%
\bibitem [{\citenamefont {Bender}(2007)}]{bender2007making}%
  \BibitemOpen
  \bibfield  {author} {\bibinfo {author} {\bibfnamefont {C.~M.}\ \bibnamefont
  {Bender}},\ }\href@noop {} {\bibfield  {journal} {\bibinfo  {journal} {Rep.
  Prog. Phys.}\ }\textbf {\bibinfo {volume} {70}},\ \bibinfo {pages} {947}
  (\bibinfo {year} {2007})}\BibitemShut {NoStop}%
\bibitem [{\citenamefont {Mostafazadeh}(2010)}]{mostafazadeh2010pseudo}%
  \BibitemOpen
  \bibfield  {author} {\bibinfo {author} {\bibfnamefont {A.}~\bibnamefont
  {Mostafazadeh}},\ }\href@noop {} {\bibfield  {journal} {\bibinfo  {journal}
  {Int. J. Geom. Methods Mod. Phys.}\ }\textbf {\bibinfo {volume} {7}},\
  \bibinfo {pages} {1191} (\bibinfo {year} {2010})}\BibitemShut {NoStop}%
\bibitem [{\citenamefont {G{\"u}nther}\ and\ \citenamefont
  {Samsonov}(2008)}]{gunther2008naimark}%
  \BibitemOpen
  \bibfield  {author} {\bibinfo {author} {\bibfnamefont {U.}~\bibnamefont
  {G{\"u}nther}}\ and\ \bibinfo {author} {\bibfnamefont {B.~F.}\ \bibnamefont
  {Samsonov}},\ }\href@noop {} {\bibfield  {journal} {\bibinfo  {journal}
  {Phys. Rev. Lett.}\ }\textbf {\bibinfo {volume} {101}},\ \bibinfo {pages}
  {230404} (\bibinfo {year} {2008})}\BibitemShut {NoStop}%
\bibitem [{\citenamefont {Kawabata}\ \emph {et~al.}(2017)\citenamefont
  {Kawabata}, \citenamefont {Ashida},\ and\ \citenamefont
  {Ueda}}]{PhysRevLett.119.190401}%
  \BibitemOpen
  \bibfield  {author} {\bibinfo {author} {\bibfnamefont {K.}~\bibnamefont
  {Kawabata}}, \bibinfo {author} {\bibfnamefont {Y.}~\bibnamefont {Ashida}}, \
  and\ \bibinfo {author} {\bibfnamefont {M.}~\bibnamefont {Ueda}},\ }\href
  {\doibase 10.1103/PhysRevLett.119.190401} {\bibfield  {journal} {\bibinfo
  {journal} {Phys. Rev. Lett.}\ }\textbf {\bibinfo {volume} {119}},\ \bibinfo
  {pages} {190401} (\bibinfo {year} {2017})}\BibitemShut {NoStop}%
\bibitem [{\citenamefont {Huang}\ \emph {et~al.}(2018)\citenamefont {Huang},
  \citenamefont {Kumar},\ and\ \citenamefont {Wu}}]{huang2018embedding}%
  \BibitemOpen
  \bibfield  {author} {\bibinfo {author} {\bibfnamefont {M.}~\bibnamefont
  {Huang}}, \bibinfo {author} {\bibfnamefont {A.}~\bibnamefont {Kumar}}, \ and\
  \bibinfo {author} {\bibfnamefont {J.}~\bibnamefont {Wu}},\ }\href@noop {}
  {\bibfield  {journal} {\bibinfo  {journal} {Physics Letters A}\ }\textbf
  {\bibinfo {volume} {382}},\ \bibinfo {pages} {2578} (\bibinfo {year}
  {2018})}\BibitemShut {NoStop}%
\bibitem [{\citenamefont {Uhlmann}(2016)}]{uhlmann16}%
  \BibitemOpen
  \bibfield  {author} {\bibinfo {author} {\bibfnamefont {A.}~\bibnamefont
  {Uhlmann}},\ }\href@noop {} {\bibfield  {journal} {\bibinfo  {journal} {Sci.
  China-Phys. Mech. Astron.}\ }\textbf {\bibinfo {volume} {59}},\ \bibinfo
  {pages} {630301} (\bibinfo {year} {2016})}\BibitemShut {NoStop}%
\bibitem [{\citenamefont {Zhang}\ \emph {et~al.}(2019)\citenamefont {Zhang},
  \citenamefont {Wang},\ and\ \citenamefont {Gong}}]{zhang2019time}%
  \BibitemOpen
  \bibfield  {author} {\bibinfo {author} {\bibfnamefont {D.-J.}\ \bibnamefont
  {Zhang}}, \bibinfo {author} {\bibfnamefont {Q.-h.}\ \bibnamefont {Wang}}, \
  and\ \bibinfo {author} {\bibfnamefont {J.}~\bibnamefont {Gong}},\ }\href@noop
  {} {\bibfield  {journal} {\bibinfo  {journal} {Physical Review A}\ }\textbf
  {\bibinfo {volume} {100}},\ \bibinfo {pages} {062121} (\bibinfo {year}
  {2019})}\BibitemShut {NoStop}%
\bibitem [{\citenamefont {Wu}\ \emph {et~al.}(2019)\citenamefont {Wu},
  \citenamefont {Liu}, \citenamefont {Geng}, \citenamefont {Song},
  \citenamefont {Ye}, \citenamefont {Duan}, \citenamefont {Rong},\ and\
  \citenamefont {Du}}]{wu2019observation}%
  \BibitemOpen
  \bibfield  {author} {\bibinfo {author} {\bibfnamefont {Y.}~\bibnamefont
  {Wu}}, \bibinfo {author} {\bibfnamefont {W.}~\bibnamefont {Liu}}, \bibinfo
  {author} {\bibfnamefont {J.}~\bibnamefont {Geng}}, \bibinfo {author}
  {\bibfnamefont {X.}~\bibnamefont {Song}}, \bibinfo {author} {\bibfnamefont
  {X.}~\bibnamefont {Ye}}, \bibinfo {author} {\bibfnamefont {C.-K.}\
  \bibnamefont {Duan}}, \bibinfo {author} {\bibfnamefont {X.}~\bibnamefont
  {Rong}}, \ and\ \bibinfo {author} {\bibfnamefont {J.}~\bibnamefont {Du}},\
  }\href@noop {} {\bibfield  {journal} {\bibinfo  {journal} {Science}\ }\textbf
  {\bibinfo {volume} {364}},\ \bibinfo {pages} {878} (\bibinfo {year}
  {2019})}\BibitemShut {NoStop}%
\bibitem [{\citenamefont {Tang}\ \emph {et~al.}(2016)\citenamefont {Tang},
  \citenamefont {Wang}, \citenamefont {Yu}, \citenamefont {He}, \citenamefont
  {Xu}, \citenamefont {Liu}, \citenamefont {Chen}, \citenamefont {Sun},
  \citenamefont {Sun}, \citenamefont {Han} \emph
  {et~al.}}]{tang2016experimental}%
  \BibitemOpen
  \bibfield  {author} {\bibinfo {author} {\bibfnamefont {J.-S.}\ \bibnamefont
  {Tang}}, \bibinfo {author} {\bibfnamefont {Y.-T.}\ \bibnamefont {Wang}},
  \bibinfo {author} {\bibfnamefont {S.}~\bibnamefont {Yu}}, \bibinfo {author}
  {\bibfnamefont {D.-Y.}\ \bibnamefont {He}}, \bibinfo {author} {\bibfnamefont
  {J.-S.}\ \bibnamefont {Xu}}, \bibinfo {author} {\bibfnamefont {B.-H.}\
  \bibnamefont {Liu}}, \bibinfo {author} {\bibfnamefont {G.}~\bibnamefont
  {Chen}}, \bibinfo {author} {\bibfnamefont {Y.-N.}\ \bibnamefont {Sun}},
  \bibinfo {author} {\bibfnamefont {K.}~\bibnamefont {Sun}}, \bibinfo {author}
  {\bibfnamefont {Y.-J.}\ \bibnamefont {Han}},  \emph {et~al.},\ }\href@noop {}
  {\bibfield  {journal} {\bibinfo  {journal} {Nat. Photonics}\ }\textbf
  {\bibinfo {volume} {10}},\ \bibinfo {pages} {642} (\bibinfo {year}
  {2016})}\BibitemShut {NoStop}%
\bibitem [{\citenamefont {Gohberg}\ \emph {et~al.}(1983)\citenamefont
  {Gohberg}, \citenamefont {Lancaster},\ and\ \citenamefont
  {Rodman}}]{gohberg1983matrices2}%
  \BibitemOpen
  \bibfield  {author} {\bibinfo {author} {\bibfnamefont {I.}~\bibnamefont
  {Gohberg}}, \bibinfo {author} {\bibfnamefont {P.}~\bibnamefont {Lancaster}},
  \ and\ \bibinfo {author} {\bibfnamefont {L.}~\bibnamefont {Rodman}},\
  }\href@noop {} {\enquote {\bibinfo {title} {Matrices and indefinite scalar
  products, volume 8 of operator theory: Advances and applications},}\ }
  (\bibinfo {year} {1983})\BibitemShut {NoStop}%
\end{thebibliography}

\end{document}